\documentclass[useAMS,usenatbib]{mnras}
\usepackage{graphicx}

\title[Inclined $v_W$ leptokurtic families]
{On the highly inclined $v_W$ leptokurtic asteroid families}
\author[V. Carruba, R. C. Domingos, S. Aljbaae, M. Huaman]{V. Carruba$^{1}$\thanks{E-mail: vcarruba@feg.unesp.br}, R. C. Domingos$^{2,1}$, S. Aljbaae$^{1}$, M. Huaman$^{1}$\\
$^{1}$UNESP, Univ. Estadual Paulista, Grupo de din\^{a}mica Orbital e
  Planetologia, Guaratinguet\'{a}, SP, 12516-410, Brazil. \\
$^{2}$UNESP, Univ. Estadual Paulista, S\~{a}o Jo\~{a}o da Boa Vista, SP, 
13874-149, Brazil.\\
}

\begin{document}

\date{Accepted 2016 August 11.  Received 2016 August 11; in original form 2016 June 29.}

\pagerange{\pageref{firstpage}--\pageref{lastpage}} \pubyear{2016}

\maketitle

\label{firstpage}

\begin{abstract}

$v_W$ leptokurtic asteroid families are families for which the distribution
of the normal component of the terminal ejection velocity field $v_W$ is 
characterized by a positive value of the ${\gamma}_2$ Pearson kurtosis, i.e.,
they have a distribution with a more concentrated peak and larger tails than 
the Gaussian one.  Currently, eight families are known to have 
${\gamma}_2(v_W) > 0.25$.  Among these, three are highly inclined asteroid
families, the Hansa, Barcelona, and Gallia families.  As observed for the
case of the Astrid family, the leptokurtic inclination distribution
seems to be caused by the interaction of these families with node secular
resonances.  In particular, the Hansa and Gallia family are crossed by 
the $s-s_V$ resonance with Vesta, that significantly alters the inclination
of some of their members.

In this work we use the time evolution of ${\gamma}_2(v_W)$ for simulated 
families under the gravitational influence of all planets and the three 
most massive bodies in the main belt to assess the dynamical importance 
(or lack of) node secular resonances with Ceres, Vesta, and Pallas for the 
considered families, and to obtain independent constraints on the family ages.
While secular resonances with massive bodies in the main belt do not 
significantly affect the dynamical evolution of the Barcelona family, they
significantly increase the ${\gamma}_2(v_W)$ values of the simulated
Hansa and Gallia families.  Current values of the ${\gamma}_2(v_W)$ for
the Gallia family are reached over the estimated family age only if
secular resonances with Vesta are accounted for.
\end{abstract}

%
%________________________________________________________________

\section{Introduction}
\label{sec: intro}

Of the three proper elements most commonly used to identify an asteroid
family, the inclination is the one usually less affected by dynamical 
evolution.  Secular resonances involving the precession frequency of
the longitude of the node of an asteroid, like the linear secular
resonance with Ceres $s-s_C$, can, however, change the inclination distribution
of families crossed by this kind of resonances \citep{Novakovic_2015}.
This is, for instance, the case of the Astrid family, that is 
characterized by a dispersion in inclination of its members at 
$a \simeq $~2.764~au much larger than that of members at other 
semi-major axis, giving this family a characteristics ``crab-like'' 
appearance in the $(a,\sin{(i)})$ plane \citep{Novakovic_2016}.  
Recently, \citet{Carruba_2016c}
used the time evolution of the Pearson kurtosis of the $v_W$ component
of terminal ejection velocities to set independent constraints on 
the Astrid family age and ejection velocity parameter $V_{EJ}$. 
Since $v_W$ can be obtained by inverting the third Gauss'equation and
is mostly dependent on  $\delta i= i-i_{ref}$, with $i_{ref}$ 
the inclination of the barycenter of the family, families whose
distribution in proper inclination is characterized by larger tails
and more concentrated peaks than that of a Gaussian distribution
would have values of Pearson kurtosis ${\gamma}_2(v_W)$ larger than
zero.  By simulating fictitious families for different values 
of ejection velocities parameter $V_{EJ}$ under the influence
of the Yarkovsky non-gravitational force, and by observing when current
values of  ${\gamma}_2(v_W)$ were reached for the Astrid families
it was possible to set independent constraints on the family 
age, $V_{EJ}$, and on key parameter determining the strength of the 
Yarkovsky force such as the mean density and surface thermal conductivity 
of family members.

Of the eight $v_W$ leptokurtic families with ${\gamma}_2(v_W) > 0.25$, 
three are highly inclined families ($\sin{(i)} > 0.3$) in the central 
main belt: the Hansa, Barcelona, and Gallia families \citep{Carruba_2016}.
It has recently been suggested that these families could be 
interacting with node secular resonances with Vesta 
\citep{Tsirvoulis_2016}.  In this work we attempt to use the numerical
tools developed for the Astrid family to i) assess the importance
(or lack of) of the $s-s_V$ secular resonance and of possible analogous
resonances with Pallas, and ii) set independent constraints on the three
families ages. Overall, we found that the use of the ${\gamma}_2(v_W)$
could indeed provide valuable hints on the importance of secular
resonances with massive bodies, and, more generally, on the whole
dynamical evolution of ${\gamma}_2(v_W)$ leptokurtic asteroid families.

\section{Asteroid families identification}
\label{sec: Fam_ide}

As a first step in our analysis of the $v_W$ leptokurtic highly inclined
families, we start by identifying the Hansa, Barcelona and Gallia family
in the space of proper elements, and by studying the local dynamics.  
For the first purpose, we use the data from \citet{Nesvorny_2015}, where
these families were identified in the domain of proper $(a,e,\sin{(i)})$
using the hierarchical clustering method and cutoff velocities of 200 m/s
for the Hansa and Gallia families and of 150 m/s for the Barcelona one.
The groups so identified have 1094 members for the Hansa family, 182 for
the Gallia group, and 306 for the Barcelona cluster.  As also discussed
in \citet{Carruba_2010}, these three families have S-type taxonomies.  
Values of the mean geometric albedos for these three groups were
0.26 for Hansa, 0.17 for Gallia, and 0.25 for Barcelona, respectively 
\citep{Nesvorny_2015}.

\citet{Carruba_2010} investigated in depth the local dynamical environment 
for these families.  The author obtained dynamical maps in the domain
of synthetic proper $(a,e)$, $(a,\sin{(i)})$, and $(e,\sin{(i)})$ domains.
Highly inclined asteroid families in the central main belt are separated
by the effect of the local secular dynamics. The strong ${\nu}_6 = g-g_6$ 
secular resonance acts as a dynamical barrier between highly inclined and low
inclined asteroids.  Of importance are also the other two main linear secular
resonances, the ${\nu}_5 = g-g_5$ and the ${\nu}_{16} = s-s_6$, that, together
with the local mean-motion resonances 3J:-1A, 8J:-3A, and 5J:-2A separate
the regions into eight different stable islands.  The region is also
crossed by several interesting non-linear secular resonance, whose detailed
identification and description can be found in \citet{Carruba_2010}.
Repeating the detailed dynamical analysis of \citet{Carruba_2010} is
of course redundant and beyond the purposes of this paper.  To allow
the reader to have a visual understanding of the complex local dynamics,
in this work we obtained dynamical maps of 7000 particles 
in the domains of synthetic proper $(a,e)$ and $(a,\sin{(i)})$ with the 
same approach described in \citet{Carruba_2010}.  We refer the reader to that
paper for a discussion of the methods and initial conditions
used for obtaining these maps.

\begin{figure*}
  \centering
  \begin{minipage}[c]{0.49\textwidth}
    \centering \includegraphics[width=3.1in]{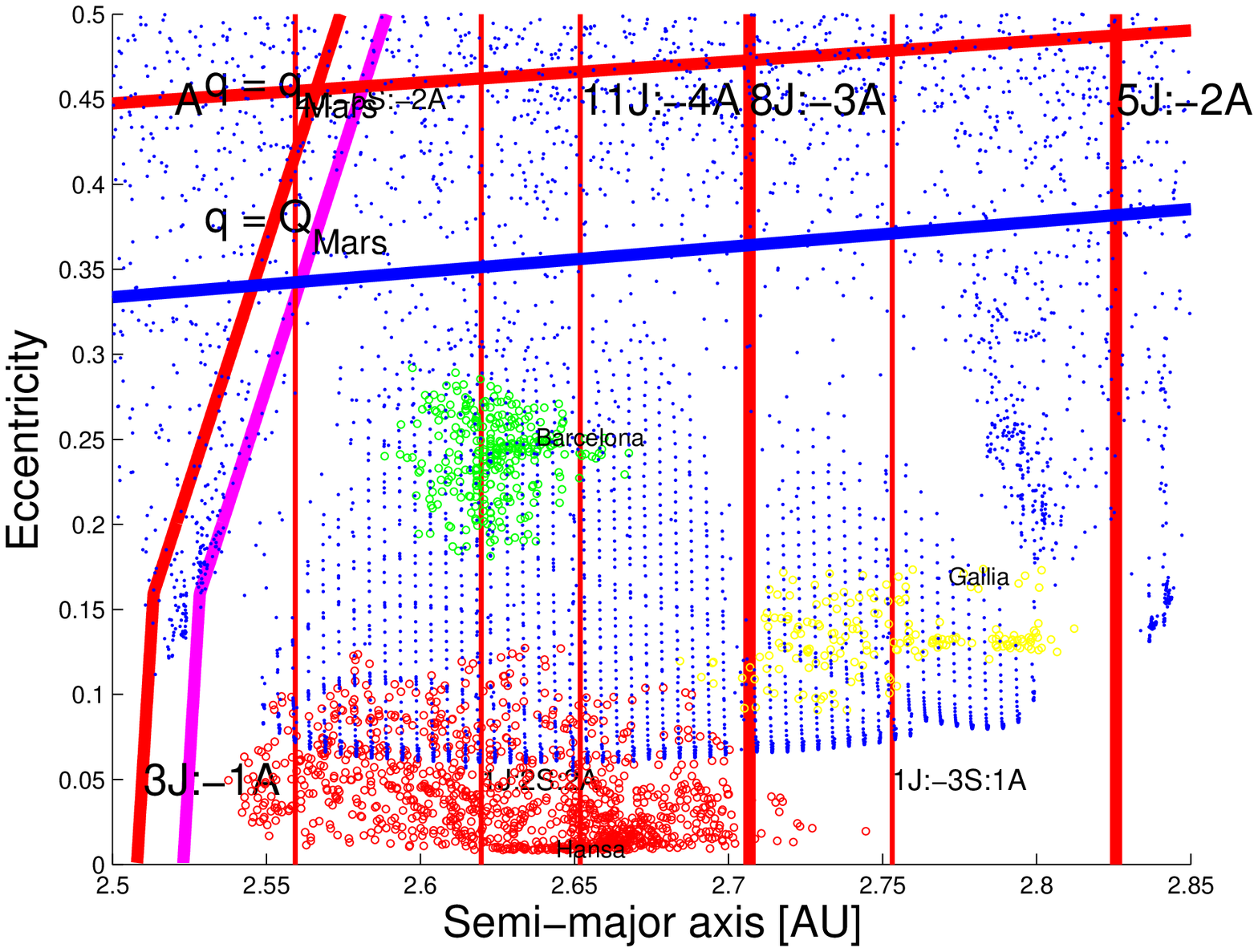}
  \end{minipage}%
  \begin{minipage}[c]{0.49\textwidth}
    \centering \includegraphics[width=3.1in]{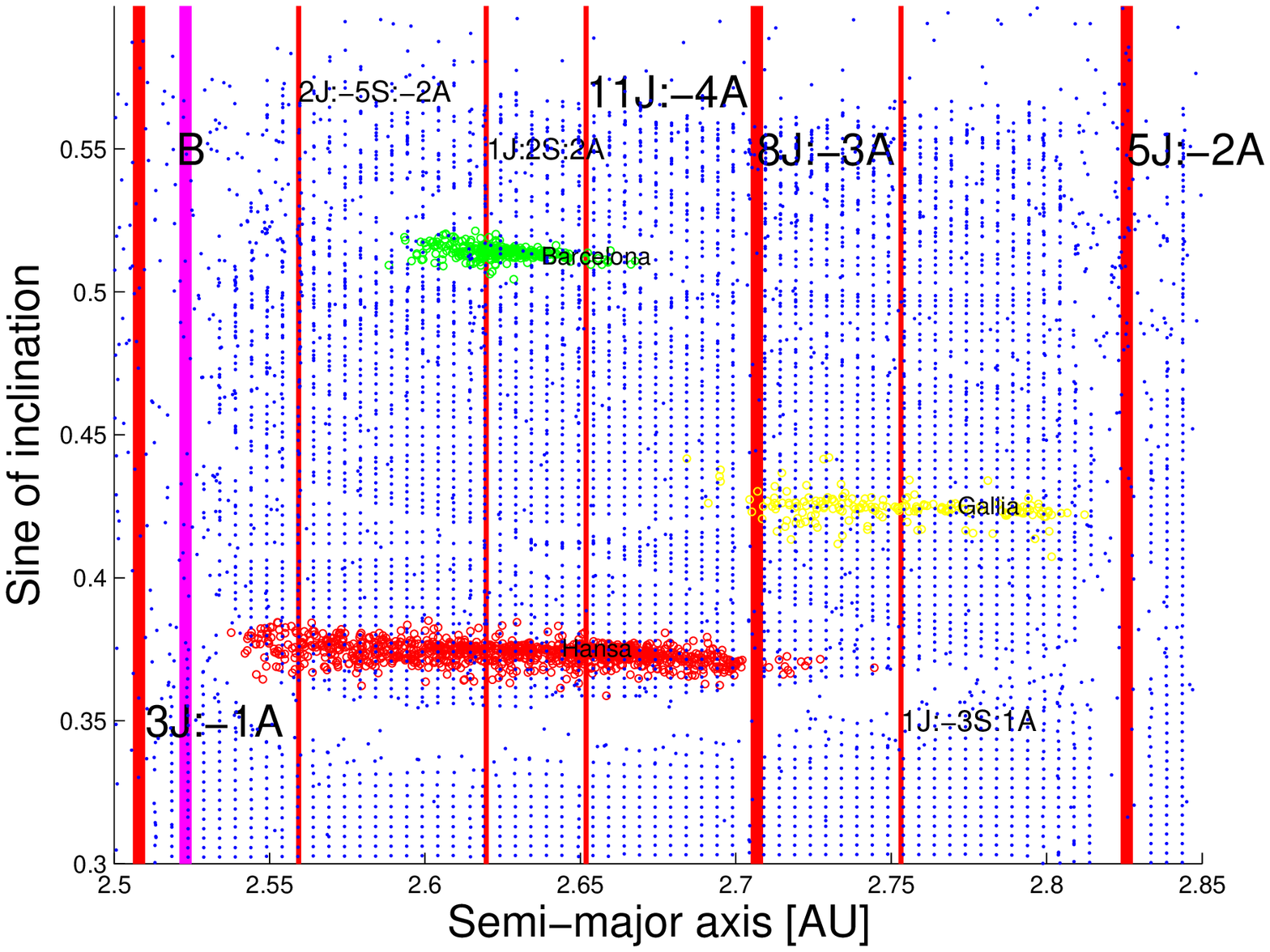}
  \end{minipage}

\caption{Dynamical maps in the proper 
$(a,e)$ (panel A) and $(a,\sin{(i)})$ (panel B) 
domains. Vertical red lines display the location of the local mean-motion
resonances. The magenta line identifies the width of the unstable chaotic 
layer near the 3J:-1A resonance, as identified in \citet{Morby_2003}.
The blue and red lines in the $(a,e)$ plane identify the orbital location
of asteroids whose pericenter $q$ is equal to the apocenter and pericenter
of Mars, respectively.  The region depleted of test particles at 
$\sin{(i)} \simeq 0.35$ in the $(a,\sin{(i)})$ plane 
is associated with orbits in librating states
of the ${\nu}_6$ secular resonance.  Other secular resonances appear as 
inclined bands of proper elements in the figure.  Red, green and yellow
full dots display the orbital location of members of the Hansa, Barcelona,
and Gallia dynamical families, respectively.} 
\label{fig: Dyn_maps}
\end{figure*}

Fig.~\ref{fig: Dyn_maps} displays our results in the $(a,e)$ (panel A) and 
$(a,\sin{(i)})$ (panel B) planes.  Vertical lines display the location
of local mean-motion resonances.  Objects with eccentricities larger than
0.35 are Mars-crossers in this region of the main belt, and are lost 
on time-scales of $\simeq$ 1 Myr.  The ${\nu}_6$ secular resonance
causes asteroids in librating states to increase their eccentricity to
Mars-crossing levels, and to become unstable.  The region associated
with this resonance appears as a strip at $\sin{(i)} \simeq 0.35$ depleted
of proper elements.  Other secular resonances appear as inclined alignments
of test particles.  Apart from the ${\nu}_5$, important for its interaction
with the Barcelona family \citep{Froeschle_1989}, and the ${\nu}_{16}$ linear
secular resonances, important non-linear secular resonances in the
region are the ${\nu}_6-{\nu}_{16}, {\nu}_5-{\nu}_{16}$, and 
$2{\nu}_6-{\nu}_5+{\nu}_{16}$ secular resonances.  Hansa, Barcelona, and
Gallia members are shown as red, green, and yellow full dots, 
respectively.  

\begin{figure}
\centering
\centering \includegraphics [width=0.49\textwidth]{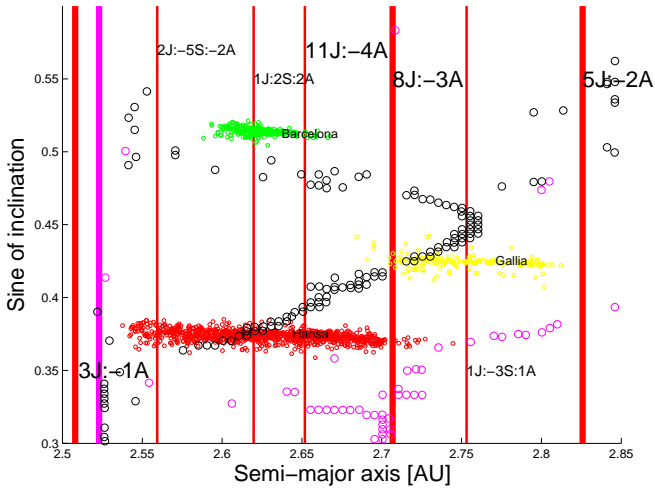}

\caption{A dynamical map in the $(a,\sin{(i)})$ domain for the same region
displayed in Fig.~\ref{fig: Dyn_maps}, but obtained also considering
the effect of Ceres, Vesta and Pallas as massive perturbers.  Black
circles display the location of likely resonators in the ${\nu}_{1V}$ 
secular resonance, while magenta circles are associated with 
likely resonators of the ${\nu}_{1P}$ resonance.  Other symbols are the
same as in Fig.~\ref{fig: Dyn_maps}, panel B.}
\label{fig: map_ai_cpv}
\end{figure}

Recently, \citet{Tsirvoulis_2016} suggested that the linear secular resonances
${\nu}_{1V}=s-s_V$ with Vesta could have played a role in the dynamical
evolution of the Barcelona and Hansa family.   To further investigate this
hypothesis, we obtained a dynamic map in the $(a,\sin{(i)})$ domain,
(node secular resonances change values of asteroids proper inclinations)
with the same initial conditions used before, but adding Ceres, Vesta, 
and Pallas as massive perturbers.  Fig.~\ref{fig: map_ai_cpv}
displays our results.  Black circles identify objects whose proper
frequency $s$ is to within $\pm 0.2$ arcsec/yr from the value 
of Vesta, assumed equal to -39.597 arcsec/yr.  These are the objects
most likely to be affected by this kind of resonant dynamics.  Magenta circles
do the same for objects whose $s$ is  $\pm 0.2$ arcsec/yr from -46.393
arcsec/yr, the proper node precession frequency for Pallas.  Pericenter
resonances with Vesta were not found to be important for this
region in \citet{Tsirvoulis_2016}, and the proper $g$ frequency of Pallas
is outside the range of values covered in this dynamical map.  For these
reasons, we did not further investigate the role of pericenter secular
resonances with massive asteroids in this work.  The ${\nu}_{1V}$ could 
in principle be affecting the dynamical evolution of the Hansa, Gallia,
and, marginally, the Barcelona families.  The ${\nu}_{1P}$ resonance,
however, does not seem to interact with any of these families, and
could play a minor role just for the case of the Hansa family.  We 
will further investigate the role played by those resonances in the
next sections.

Having briefly revised the local dynamics, we then turn our attention 
to the taxonomic properties of the three studied families.  Taxonomic
properties and geometric albedo values of all highly inclined families
in the central main belt were studied in \citet{Carruba_2010} in some
detail.  The three families all belong to the S-type taxonomic class.
There are 94 objects with photometric data compatible with a S-type 
composition in the Sloan Digital Sky Survey-Moving Object Catalog data, 
fourth release (SDSS-MOC4 hereafter, \citet{Ivezic_2001}) in this region.
470 objects have geometric albedo and absolute magnitude information 
in the WISE and NEOWISE database \citep{Masiero_2012}.

\begin{figure*}
  \centering
  \begin{minipage}[c]{0.49\textwidth}
    \centering \includegraphics[width=3.1in]{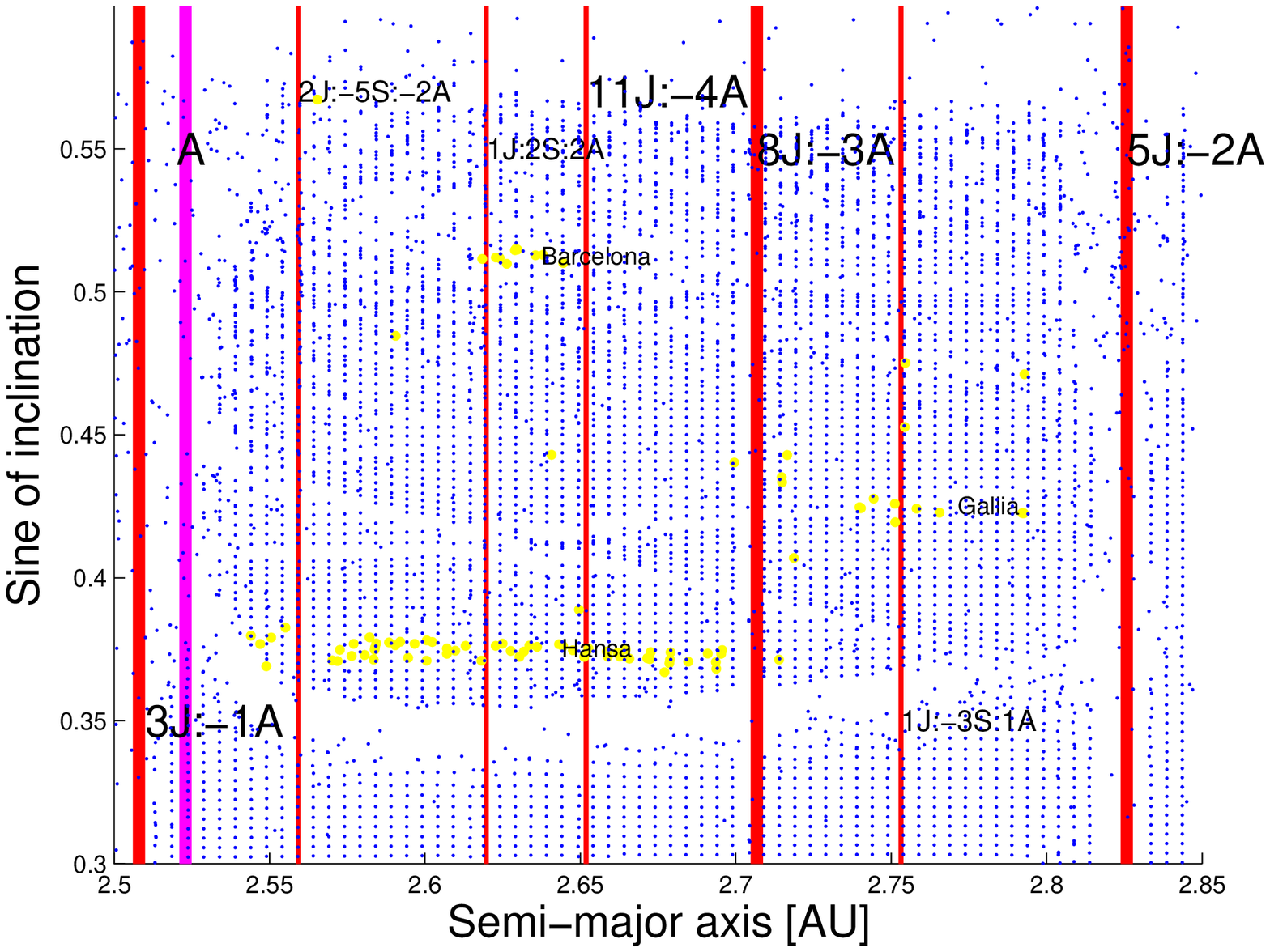}
  \end{minipage}%
  \begin{minipage}[c]{0.49\textwidth}
    \centering \includegraphics[width=3.1in]{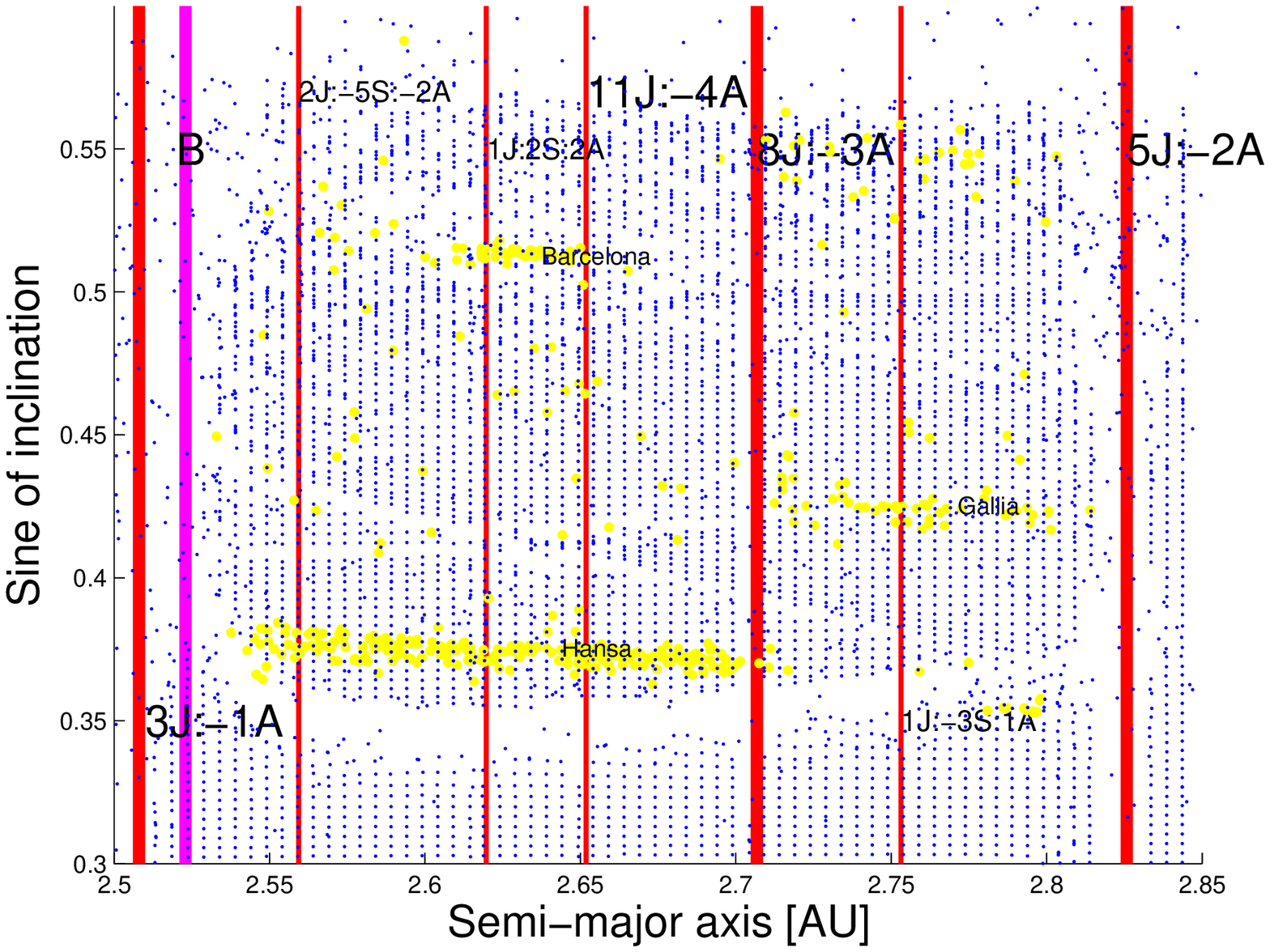}
  \end{minipage}

\caption{Panel A: yellow full dots identify objects whose SDSS-MOC4 
photometric data is compatible with an S-type composition.  Panel B:
objects with WISE albedo $p_V$ in the range from 0.15 to 0.30.  
Other symbols are the same as in Fig.~\ref{fig: Dyn_maps}, panel B.} 
\label{fig: GHB_tax}
\end{figure*}

Fig.~\ref{fig: GHB_tax} display asteroids whose SDSS-MOC4 photometric
data is compatible with a S-type composition, according to the 
classifications method of \citet{DeMeo_2013} (panel A).  Panel B
displays objects whose WISE geometric albedo $p_V$ has values compatible
with a S-complex taxonomy, (i.e., $0.12 < p_V < 0.30$,
\citet{Masiero_2012}).  As can be seen from the figure, most of the 
objects with photometric and albedo S-type compatible data are indeed 
associated with the three studied families.   Using the SDSS-MOC4 data,
we tried to obtain halos for the three families with the method
discussed in \citet{Carruba_2016b} for the Koronis family.
In this method, asteroids with SDSS-MOC4 data are considered to be members
of the halo of the family if their values of proper eccentricity and 
inclination are in a range from the center of dynamical family to within 
four standard deviations of $e$ and $\sin{(i)}$ of the distribution observed 
for the HCM family.  We applied this method for the three studied families,
but, unfortunately, we are limited by small number statistics for the
cases of the Gallia and Barcelona families, that have halos of less than 10
members.  The Hansa family has a halo of 63 members, but its distribution
in proper inclination is comparable to that of the HCM family.  For the
purpose of studying the Kurtosis of the $v_W$ component of terminal ejection
velocities, we are therefore left using standard HCM data alone.
The apparent lack of significant halos for these three families
may be caused by the fact that these groups are contained in stable islands
surrounded by unstable regions, which limits the number of long-term
surviving outlying asteroids.

\citet{Carruba_2016} studied the shape of the $v_W$ component of the ejection
velocity field of these three families, that are among the most leptokurtic
among the studied group.  For the sake of the reader not familiar with that 
work, we summarize in Table~\ref{table: families_kurt} the result of that 
study.  The first two column report the Family Identification Number (FIN),
as defined in \citet{Nesvorny_2015}, and the family identification and
name.  The third and fourth columns report the values of ${\gamma}_2(v_W)$
for the whole family and for the $D_3$ population with $2.5 < D < 3.5$ km,
respectively.  The fifth column displays the result of the 
Jarque-Bera statistical test of the distribution being compatible
with a Gaussian distribution (0.5\% being the null probability level
of the two distributions being compatible).  Finally, the 
sixth column reports the estimated age and its error from 
\citet{Nesvorny_2015}, computed using the procedure
discussed in \citet{Carruba_2016}, Sect. 4.  
The time evolution of the $v_W$ component of 
simulated families will be discussed later on in this paper.

\begin{table}
\begin{center}
\caption{Values of ${\gamma}_2(v_W)$ of the whole family 
(3rd column), the $2.0 < D < 4.0$~km ($D_3$) members (4rd column), 
the $p$ coefficient of the jbtest (5th column), and estimated family age
with its error (6th column) from \citet{Nesvorny_2015} for the Hansa, 
Barcelona,and Gallia families.}
\label{table: families_kurt}
\begin{tabular}{|c|c|c|c|c|c|}
\hline
      &        &                   &                  &            &        \\ 
FIN   & Family & ${\gamma}_2(v_W)$ & ${\gamma}_2(v_W)$ & $p_{jbtest}$ &   Age  \\
      & Name   &      All          & $D_3$             &   (\%)     &   [Myr]\\
      &        &                   &                  &            &        \\ 
\hline 
      &        &                   &                  &            &        \\ 

803 & 480 Hansa     & 0.81 & 1.17 & 0.1 & $2430\pm600$ \\
805 & 945 Barcelona & 1.48 & 1.32 & 0.5 & $250\pm10$    \\
802 & 148 Gallia    & 2.02 & 3.39 & 0.1 & $650\pm60$    \\
      &        &                   &                  &            &        \\ 
\hline
\end{tabular}
\end{center}
\end{table}

\section{Chronology}
\label{sec: Chron}

There is a considerable range of possible values for the age of the 
Hansa family in the literature.  \citet{Carruba_2010}, using the 
method of Yarkovsky isolines, provided an upper limit for 
the family age of 1600 Myr old (see also \citet{Broz_2013}.  
The estimate from \citet{Nesvorny_2015} was 
of 2430$\pm$60 Myr, while \citet{Spoto_2015}, using a V-shape criteria, 
assessed the family age to be in the range 420-1170 Myr. 
The age of the Barcelona family was estimated by  \citet{Carruba_2010}
(see also \citet{Broz_2013}) to have an upper limit of 350 Myr 
and to be in the range of 250$\pm$10 Myr by \citet{Nesvorny_2015}.  
The Gallia family had an upper limit of 450 Myr in \citet{Carruba_2010} 
(see also \citet{Broz_2013}) and an age estimate of 650$\pm$60 Myr in 
\citet{Nesvorny_2015}.  No estimates for the ages of the Barcelona and 
Gallia family was provided in \citet{Spoto_2015}.

\begin{figure*}
  \centering

  \begin{minipage}[c]{0.67\textwidth}
    \centering \includegraphics[width=0.67\textwidth]{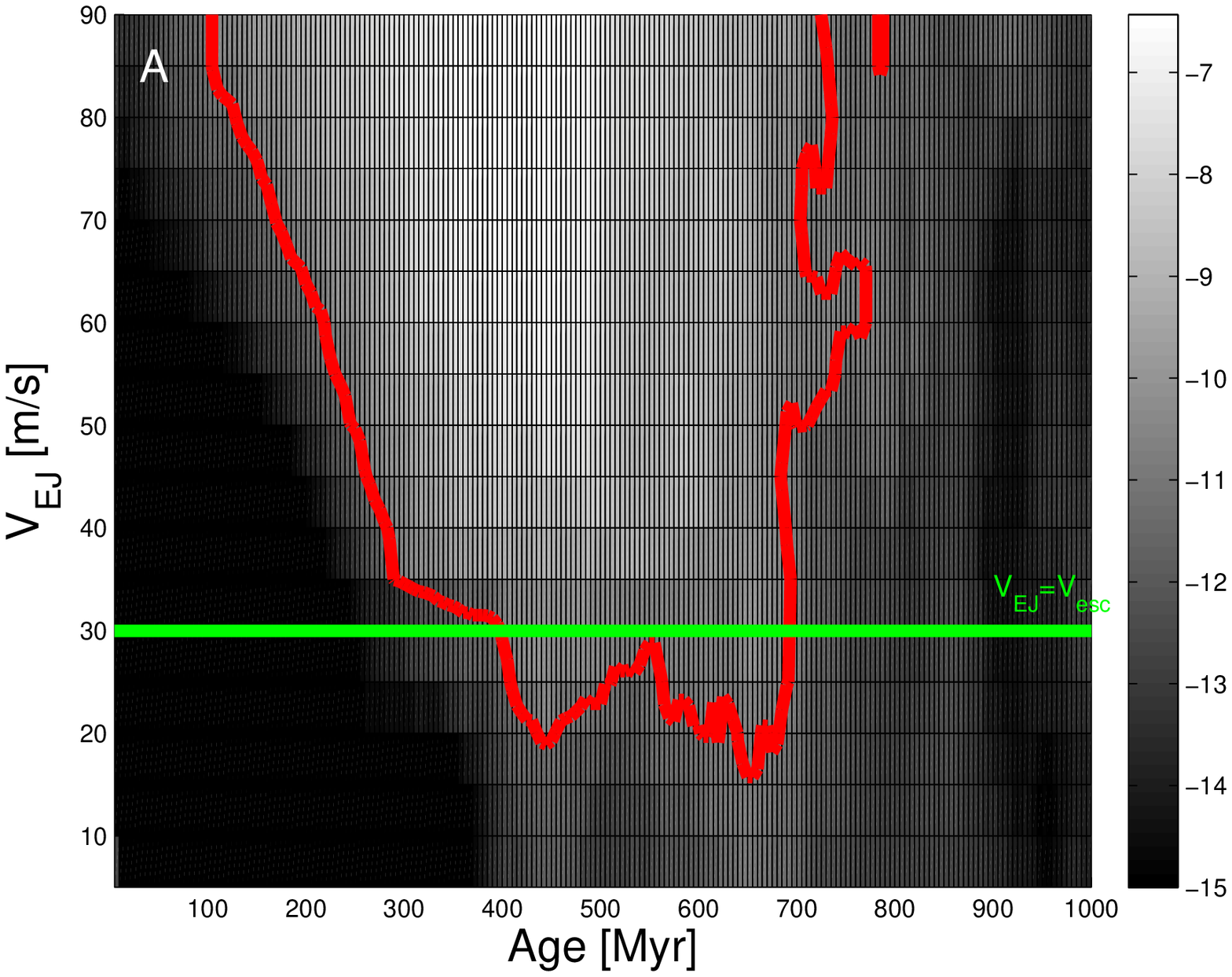}
  \end{minipage}
  \begin{minipage}[c]{0.49\textwidth}
    \centering \includegraphics[width=3.1in]{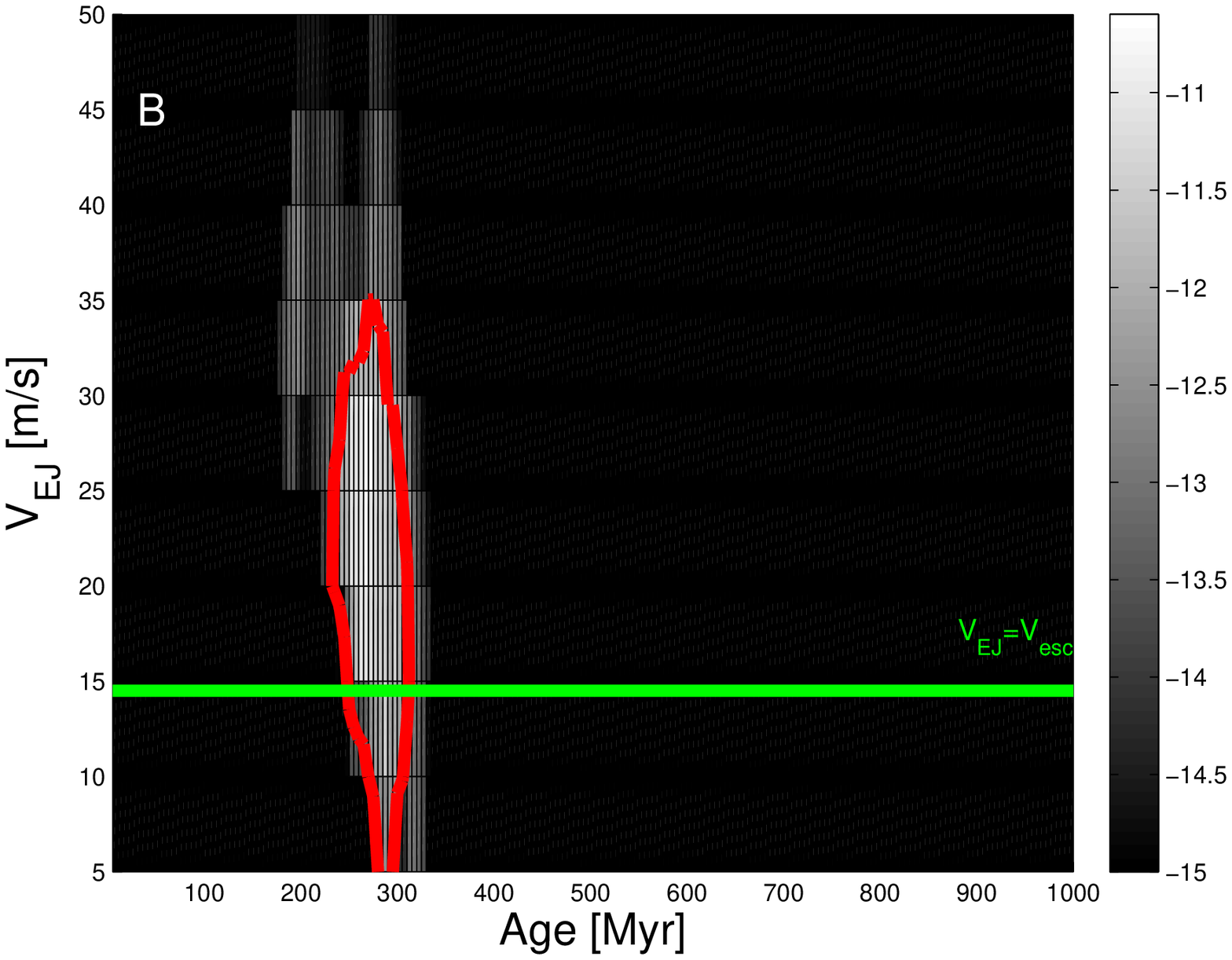}
  \end{minipage}%
  \begin{minipage}[c]{0.49\textwidth}
    \centering \includegraphics[width=3.1in]{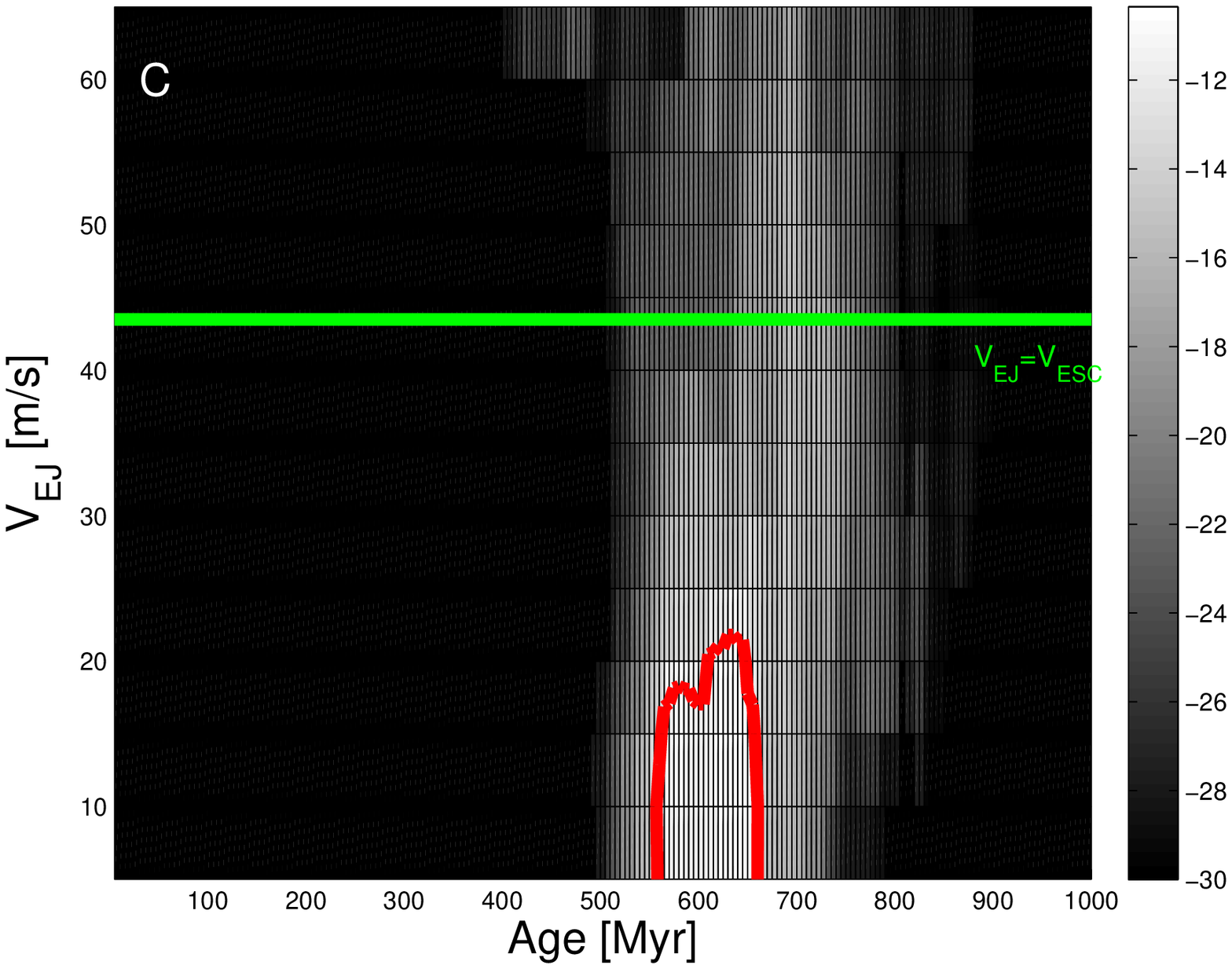}
  \end{minipage}
\caption{Target function ${\psi}_{\Delta C}$ values in ($Age,V_{EJ}$) 
plane for the Hansa (panel A), Barcelona (panel B), and Gallia families
(panel C).  The horizontal green lines display the value of 
the estimated escape velocities $V_{ESC}$ from the parent body.  The red
lines display the contour level of ${\psi}_{\Delta C}$ associated
with a 1-sigma probability that the simulated and 
real distribution were compatible.}
\label{fig: cont_psi_ghb}
\end{figure*}

Here we try to obtain a new estimate of the families ages using the 
approach described in \citet{Carruba_2015a},
that uses a Monte Carlo method \citep{Vokrouhlicky_2006a,
Vokrouhlicky_2006b, Vokrouhlicky_2006c, Novakovic_2010}.
The method has been described in several previous papers, so here
we just shortly summarized the approach.  Interested readers can 
found more details in \citet{Carruba_2015a}.
Basically, fictitious families with different values of $V_{EJ}$, a 
parameter describing the shape of the family ejection velocity field,
are generated and then evolved under the influence of the Yarkovsky
and YORP effects, and taking into account that solar luminosity
was less intense in the past.  The obtained  
distribution of a $C$ parameter, that depends on the asteroids 
semi-major axis and absolute magnitude, is then compared to the one 
observed for the real asteroid family, and a ${\chi}^2$-like variable  
${\psi}_{\Delta C}$ is then used to evaluate which fictitious family 
best approximate the $C$ distribution of the real asteroid group.  
We applied this method to the Hansa, Barcelona, and Gallia families, and 
Fig.~\ref{fig: cont_psi_ghb} displays our results in the ($Age,V_{EJ}$) plane.  
The radius of the family parent body, as estimated from \citet{Nesvorny_2015},
and  the escape velocity $V_{ESC}$ are reported in Table~\ref{table: ghb_ages}. 
Since \citet{Carruba_2016} showed that most asteroid families have values 
of $V_{EJ}$ not greater than 1.5 $V_{ESC}$, we considered 
values of $V_{EJ}$ going from 0 up to 90 m/s, i.e., equal to 1.5 
the estimated escape velocity ($\simeq 60$~m/s) from the Hansa
parent body, the body with the largest escape velocity among the families
here studied.

At a 1-sigma level of probability of the simulated family distribution
being compatible with the real one (red curve in Fig.~\ref{fig: cont_psi_ghb},
associated ${\psi}_{\Delta C}=10.73$ and 12 degree of freedoms for our 
distribution) we found that $T = 460^{+280}_{-360}$ Myr, and 
$V_{EJ}= 80^{+10}_{-65}$ m/s for the Hansa family, $T = 265^{+45}_{-35}$ Myr, and 
$V_{EJ}= 15^{+20}_{-15}$ m/s for the Barcelona family, and $T = 630^{+30}_{-70}$ 
Myr, and $V_{EJ}= 5^{+15}_{-5}$ m/s for the Gallia one.  Again, our results 
are summarized in Table~\ref{table: ghb_ages}.

\begin{table*}
\begin{center}
\caption{Number of family members ($N_{mem}$), mean geometric albedo
($p_V$), estimated radius of the parent body ($R_{PB}$), escape velocities 
($V_{ESC}$), estimated $V_{EJ}$, and ages $T$ for the Hansa, Barcelona, and 
Gallia families.}
\label{table: ghb_ages}
\vspace{0.5cm}
\begin{tabular}{|c|c|c|c|c|c|c|}
\hline
          &            &      &           &         &          &        \\
Family    &   $N_{mem}$ & $p_V$ &  $R_{PB}$ & $V_{ESC}$ & $V_{EJ}$  &  $T$   \\
          &            &      &  [km]    &   [m/s]  & [m/s]    &  [Myr]  \\
          &            &      &          &          &          &        \\
\hline
          &      &      &        &       &               &                  \\
Hansa     & 1094 & 0.26 &   28.0 &  30.0 & 80$^{+10}_{-65}$ & 460$^{+280}_{-360}$ \\
Barcelona &  306 & 0.25 &   13.5 &  14.5 & 15$^{+20}_{-15}$ & 265$^{+45}_{-35}$  \\
Gallia    &  182 & 0.17 &   40.5 &  43.4 &  5$^{+17}_{-5}$  & 630$^{+30}_{-70}$  \\
          &      &      &        &       &               &                 \\
\hline
\end{tabular}
\end{center}
\end{table*}

\section{Dynamical evolution of $v_W$ leptokurtic families}
\label{sec: Dyn_evol}

The time evolution of the kurtosis of the $v_W$ component of the ejection
velocity field was recently used by \citet{Carruba_2016c} to set constraints
on the age and acceptable values of key parameters describing the Yarkovsky
force, such as the surface thermal conductivity and asteroid density of
the Astrid asteroid family.   Here we use the same approach to investigate
the dynamics of the three $v_W$ leptokurtic highly inclined families.
The set-up of the simulations was discussed in \citet{Carruba_2016c},
interested readers could find more details in that paper.
Basically, we simulated fictitious families with their currently
observed size-frequency distribution, values of the parameters affecting
the strength of the Yarkovsky force typical of S-type asteroids according to 
\citet{Broz_2013}, i.e., bulk and surface density, ${\rho}_{bulk}$ 
and ${\rho}_{surf}$, equal to$ 1500$ and $2500$~kg/m$^3$, respectively, 
thermal conductivity $K =0.001$~W/m/K, thermal capacity equal to 
$C_{th} = 680$~J/kg/K, Bond albedo $A_{Bond} =0.1$\footnote{Since the mean
geometric albedo value of the Gallia family is lower than that
of the other two families, and since this could imply a lower
value of the Bond albedo, we also performed two additional
sets of simulations for this family with a value of Bond albedo 
$A_{Bond} =0.07$.  The overall trend of the results of this simulations
was compatible with that of the standard simulations with $A_{Bond} =0.1$.},  
and infrared emissivity $\epsilon = 0.9$.
We also generated fictitious families with the optimal values
of the ejection parameter $V_{EJ}$ found in Sect.~\ref{sec: Chron}, for the 
three families  Particles were integrated with
$SWIFT\_RMVSY$, the symplectic integrator developed by \citet{Broz_1999}
that simulates the diurnal and seasonal versions of the Yarkovsky effect, 
over 1000 Myr for the Hansa family, 600 Myr for the Barcelona group,
and 800 Myr for the Gallia cluster, a time long enough to cover the 
putative estimated ages of these families.  Two sets of simulations were
performed. In the first we accounted for all planets from Mercury to 
Neptune, while in the second we also include Ceres, Vesta, and Pallas
as massive bodies. Once proper elements were obtained, then values 
of $v_W$ were computed by inverting the third Gauss equation
\citep{Murray_1999}:

\begin{equation}
\delta i = \frac{(1-e^2)^{1/2}}{na} \frac{cos(\omega+f)}{1+e cos(f)} \delta v_W. 
\label{eq: gauss_3}
\end{equation}

where $\delta i= i-i_{ref}$, with $i_{ref}$ the inclination of the barycenter
of the family, and $f$ and $\omega+f$ assumed equal to 30$^{\circ}$ 
and 50.5$^{\circ}$, respectively. As discussed in \citet{Carruba_2016c}, 
the shape of the $v_W$ distribution (and therefore its kurtosis), 
are not strongly dependent on the values of $f$ and $\omega+f$.

\begin{figure*}
  \centering

  \begin{minipage}[c]{0.49\textwidth}
    \centering \includegraphics[width=3.1in]{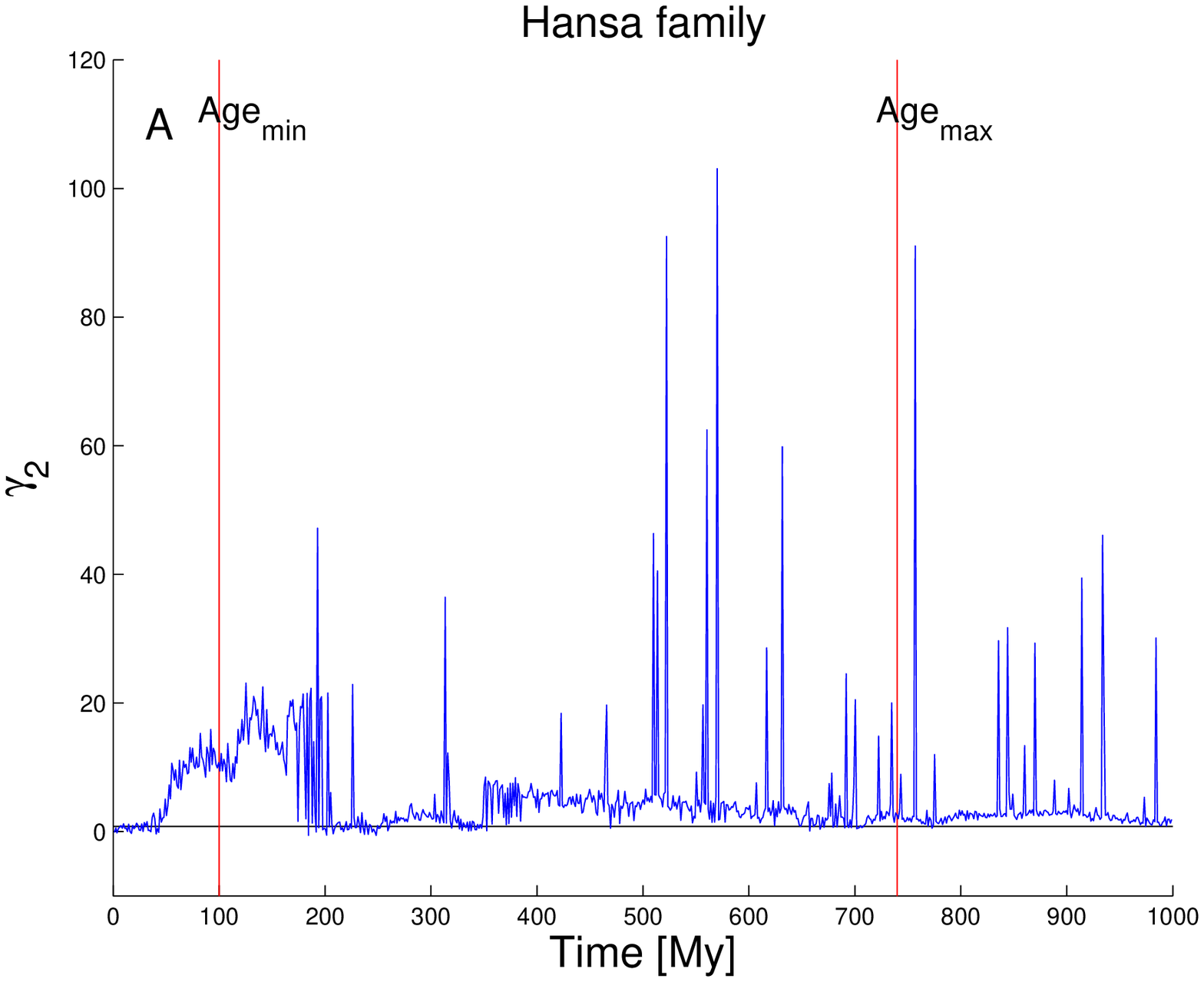}
  \end{minipage}%
  \begin{minipage}[c]{0.49\textwidth}
    \centering \includegraphics[width=3.1in]{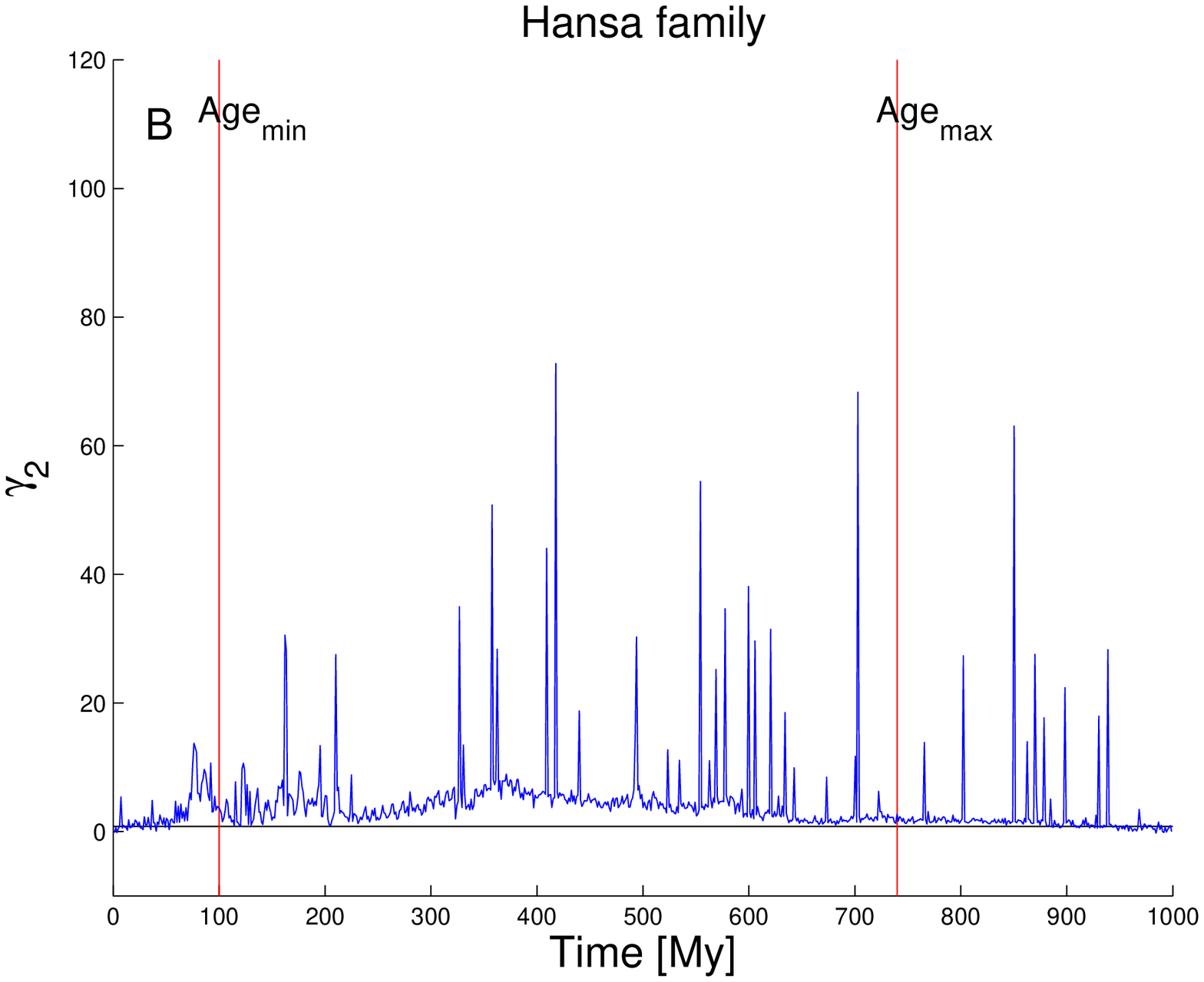}
  \end{minipage}

  \begin{minipage}[c]{0.49\textwidth}
    \centering \includegraphics[width=3.1in]{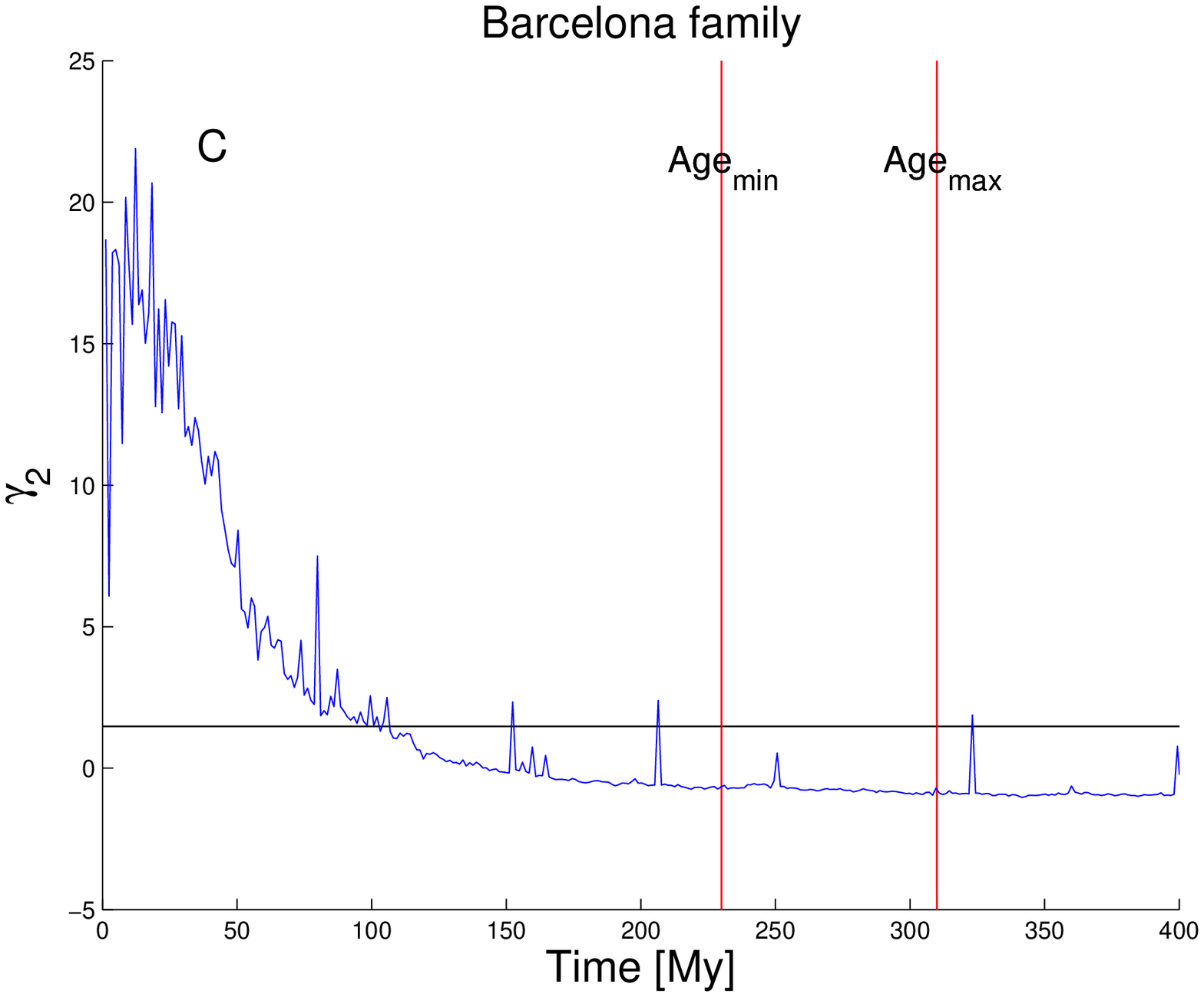}
  \end{minipage}%
  \begin{minipage}[c]{0.49\textwidth}
    \centering \includegraphics[width=3.1in]{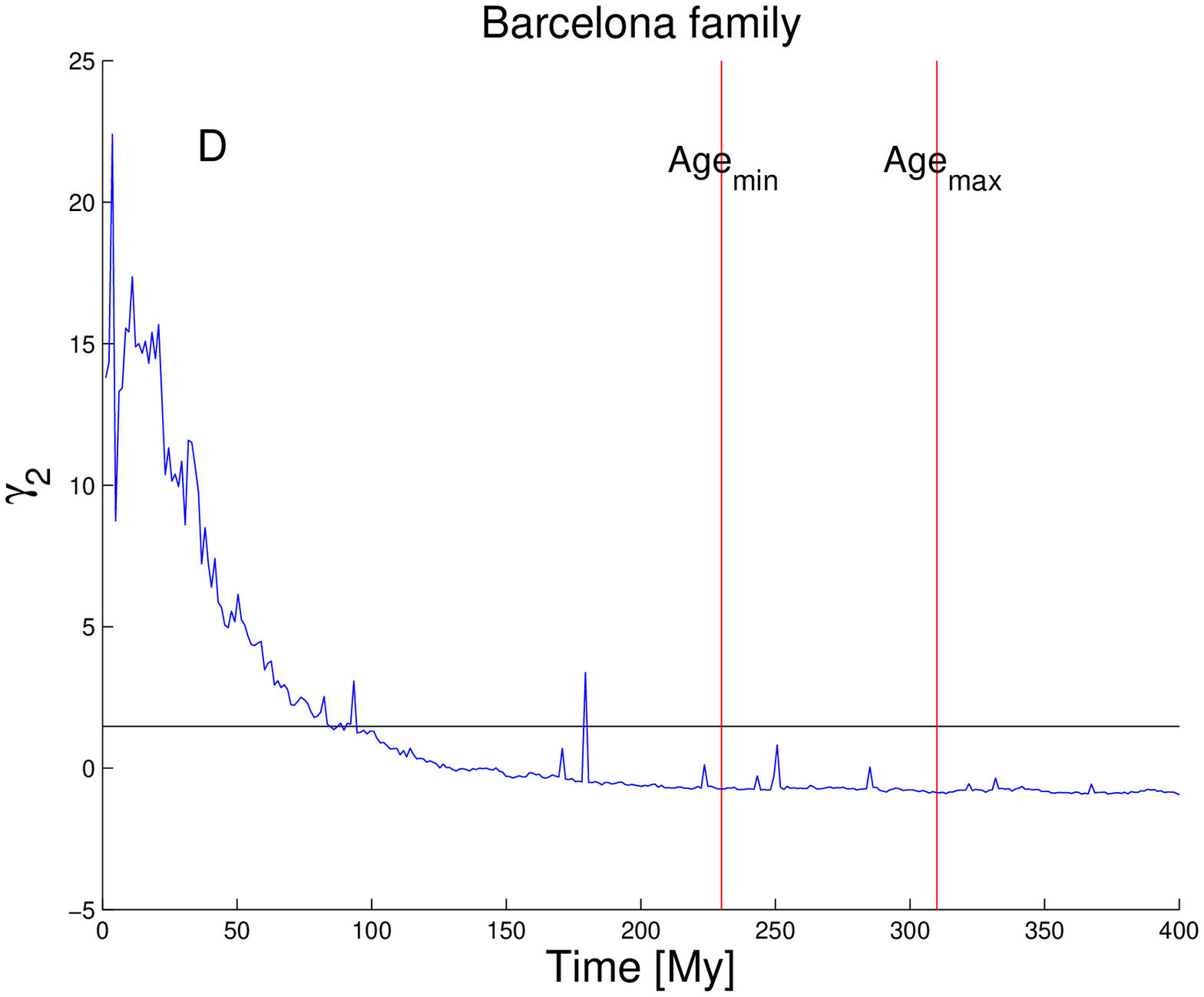}
  \end{minipage}

  \begin{minipage}[c]{0.49\textwidth}
    \centering \includegraphics[width=3.1in]{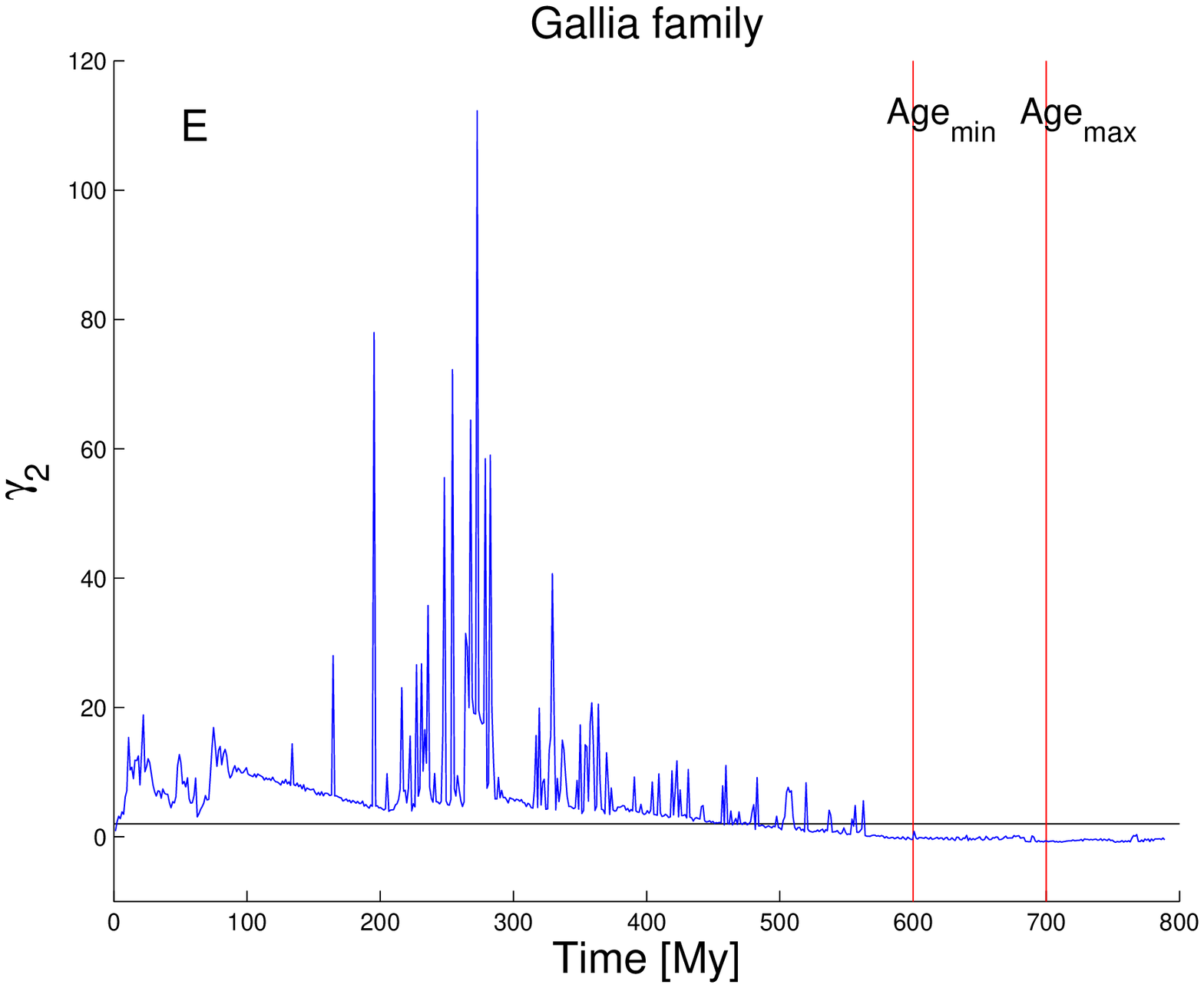}
  \end{minipage}%
  \begin{minipage}[c]{0.49\textwidth}
    \centering \includegraphics[width=3.1in]{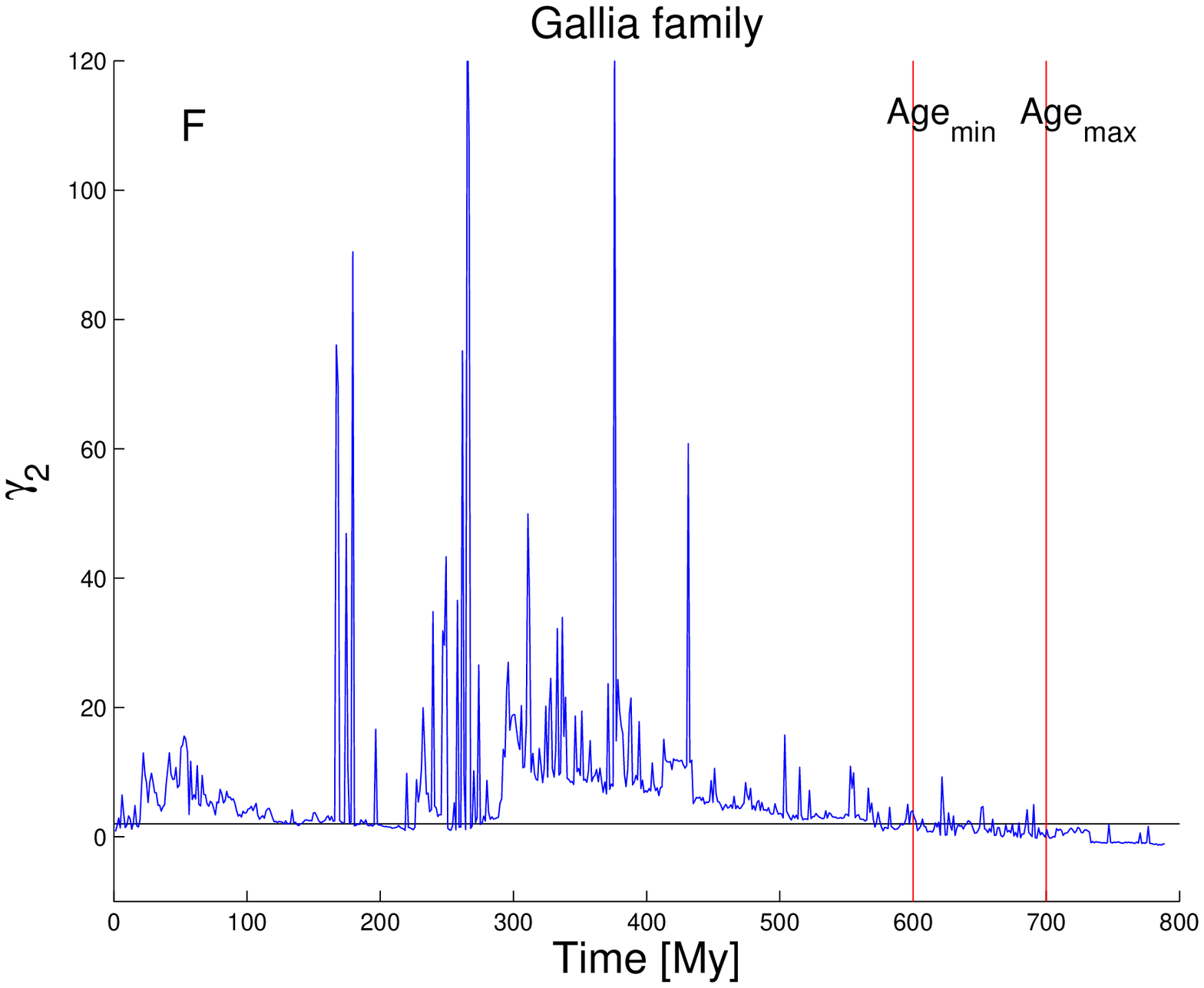}
  \end{minipage}

\caption{Time dependence of the kurtosis of the $v_W$ component of the 
ejection velocity field (${\gamma}_2{(v_W)}$) for the simulated Hansa, 
Barcelona, and Gallia families (panels A, C, and E, respectively).  Panels
B, D, and E display the same, but for the simulation where Ceres, Pallas,
and Vesta were considered as massive perturbers.  The horizontal black
line display the current values of the families ${\gamma}_2{(v_W)}$,
while the vertical red lines show the estimated range of possible ages.} 
\label{fig: gamma_vW}
\end{figure*}

Fig.~\ref{fig: gamma_vW} displays our results for the set of six simulations.
Since the value of ${\gamma}_2{(v_W)}$ increased significantly when 
isolated objects drift beyond 4 sigma values in $\sin{(i)}$ from the
center of the family, as in \citep{Carruba_2016} we eliminated from
our computations of this parameters objects with inclination beyond
that range.  In particular, this meant considering asteroids
with inclination between $21.6^{\circ}$ and $22.4^{\circ}$ for the Hansa
family, between $27.8^{\circ}$ and $29.1^{\circ}$ for the Barcelona
family, and between $24.0^{\circ}$ and $26.2^{\circ}$ for the Gallia
family.   Results for the Hansa family without the effect of massive 
asteroids show
that values of ${\gamma}_2{(v_W)}$ are compatible with the current one 
for times larger than 200 Myr, which sets a lower limit on the family age.
Including the massive asteroids only slightly alters this scenario, which
suggest that the effect of resonances with Vesta for the Hansa family should
be minor, when compared with other local resonances able to affect 
the inclination in the region.  Spikes in the time behavior
of ${\gamma}_2{(v_W)}$ are associated with isolated asteroids whose
inclination temporarily approached values of $\sin{(i)}$ close to the
limits considered in our analysis. Of course, changing the allowed limits
of $\sin{(i)}$ could modify the length and the shape of the isolated spikes
observed in Fig. 5.  Since our goal in this paper is, however, to assess
the importance of different dynamical models and since we are using
the same limits for the model with and without massive asteroids,
we believe that our approach should be reasonable.

Concerning the Barcelona family, resonances with massive asteroids are not 
important for this family, as shown in Fig.~\ref{fig: gamma_vW}, panels C and
D. Results are essentially identical with and without massive asteroids,  as
expected from the results of Sect.~\ref{sec: Fam_ide}, that showed that 
the Barcelona family is not actually crossed by the ${\nu}_{1V}$ resonance, 
or other resonances with massive bodies.  This negative result, however,
confirm the usefulness of the ${\gamma}_2{(v_W)}$ as a tool to investigate
the long-term behavior of secular dynamics.  As observed for the Astrid
family \citep{Carruba_2016c}, commonly used values of the key parameters
density and thermal conductivity for S-type families are not able 
to produce the currently observed value of ${\gamma}_2{(v_W)}$ over
the estimated age of the family.  This could be either caused
by i) the fact that the mean values of density and thermal conductivity
for members of the Barcelona family could be higher, or ii) that
the actual age of this family could be younger than what obtained
from estimates from the Monte Carlo method of section~\ref{sec: Chron}.
An analysis of the full dependence of the ${\gamma}_2{(v_W)}$ time behavior
on values of density and thermal conductivity for the Barcelona family
performed in the same way as recently done for the Astrid cluster seems
to be outside the goals of this paper, that focused on studying the 
effectiveness of the use of  ${\gamma}_2{(v_W)}$ as a tool to investigate
the long-term effect of secular dynamics.  But it certainly remains an 
interesting topic for future research.

Finally, the case of the Gallia family is of particular interest.  If 
we do not consider the effect of secular resonances with Vesta, 
the simulated ${\gamma}_2{(v_W)}$ does not reach current values over the 
estimated age of the family.  But there is an excellent agreement
if we include secular perturbations from massive asteroids, as shown 
in Fig.~\ref{fig: gamma_vW}, panel F.  The larger values of 
${\gamma}_2{(v_W)}$ obtained when massive asteroids are considered and
the agreement with estimates of the Gallia family age obtained with 
independent methods represent, in our opinion, some of the newest
and most interesting results of this work.

\section{Conclusions}
\label{sec: Conc}

Our results could be summarized as it follows:

\begin{enumerate}

\item We identified the Hansa, Barcelona, and Gallia families in the
domain of proper elements, obtained dynamical maps in the domains of
proper $(a,e)$ and $(a,\sin{(i)})$, and mapped the location in 
the $(a,\sin{(i)})$ of the three node resonance with Ceres, Vesta,
and Pallas.  The Hansa and Gallia families are crossed by the ${\nu}_{1V} = 
s-s_V$ secular resonances.  All families are S-complex group, all characterized
by relatively large values of ${\gamma}_2{(v_W)}$.

\item We obtained age and terminal ejection velocities 
estimates for the three families using a Monte Carlo
method to simulate the Yarkovsky and stochastic YORP evolution in proper
$a$ of family members.  At one sigma confidence level we found that 
$T = 460^{+280}_{-360}$ Myr, and $V_{EJ}= 80^{+10}_{-65}$ m/s for the Hansa family, 
$T = 265^{+45}_{-35}$ Myr, and $V_{EJ}= 15^{+20}_{-15}$ m/s for the Barcelona 
family, and $T = 630^{+30}_{-70}$ Myr, and $V_{EJ}= 5^{+15}_{-5}$ m/s for 
the Gallia one.  Our results are summarized in Table~\ref{table: ghb_ages}.

\item Simulated the dynamical evolution of fictitious families with
values of the ejection velocity parameters $V_{EJ}$ obtained from our previous
analysis under the gravitational influence of all planets, the Yarkovsky
force, and the effect of Ceres, Vesta, and Pallas.  The ${\gamma}_2{(v_W)}$
parameter was computed as a function of time for all simulated family
members, and we monitored when its value was comparable to the currently
observed ones.  The Gallia and, in a less measure, the Hansa 
families were significantly affected
by secular resonances with Vesta.  Current values of ${\gamma}_2{(v_W)}$
for the Gallia family could only be reached over the estimated
family age if secular resonances with Vesta were accounted for.
Conversely, secular resonances with main belt massive bodies play no
significant role in the evolution of the Barcelona family.  Independent
constraints on the family age can be set by the time behavior of
the ${\gamma}_2{(v_W)}$ parameter.

\end{enumerate}

Overall, we found that the ${\gamma}_2{(v_W)}$ parameter could be an invaluable
tool for providing hints about the relative importance of secular dynamics,
and to set constraints on the ages, ejection velocity fields, and 
key parameters influencing the Yarkovsky force, such as the mean 
density and surface conductivity of $v_W$ leptokurtic families, and could 
be in principle applied to other similar families identified in 
\citet{Carruba_2016}.

\section*{Acknowledgments}
We are grateful to the reviewer of this paper, Dr. Bojan Novakovi\'{c},
for comments and suggestions that greatly improved the quality of this
paper. We would like to thank the S\~{a}o Paulo State Science Foundation 
(FAPESP) that supported this work via the grant 2016/04476-8, and the
Brazilian National Research Council (CNPq, grant 305453/2011-4).
This publication makes use of data products from the Wide-field 
Infrared Survey Explorer (WISE) and NEOWISE, which are a joint project 
of the University of California, Los Angeles, and the Jet Propulsion 
Laboratory/California Institute of Technology, funded by the National 
Aeronautics and Space Administration.

\bsp

\label{lastpage}

\begin{thebibliography}{}

\bibitem[Bro\v{z}(1999)]{Broz_1999} Bro\v{z}, M., 1999.
Thesis, Charles Univ., Prague, Czech Republic.

\bibitem[Bro\v{z} et al.(2013)]{Broz_2013} Bro\v{z}, M., Morbidelli, A., 
 Bottke, W.~F., et~al. 2013, A\&A, 551, A117

\bibitem[Carruba(2010)]{Carruba_2010} Carruba, V. 2010. MNRAS, 408, 580.

\bibitem[Carruba et al.(2015)]{Carruba_2015a} Carruba, V., Nesvorn\'{y}, D., 
Aljbaae, S., Domingos, R. C., Huaman, M. E., 2015, MNRAS, 451, 4763.

\bibitem[Carruba \& Nesvorn\'{y}(2016)]{Carruba_2016} Carruba, V., 
Nesvorn\'{y}, D., 2016, MNRAS, 457, 1332.
  
\bibitem[Carruba et al.(2016)]{Carruba_2016b} Carruba, V., 
Nesvorn\'{y}, D., Aljbaae, S., 2016, Icarus, 271, 57. 

\bibitem[Carruba (2016)]{Carruba_2016c} Carruba, V., MNRAS, 2016, 
461, 1605.

\bibitem[DeMeo \& Carry(2013)]{DeMeo_2013} DeMeo, F., 
Carry, B., 2013, Icarus, 226, 723.

\bibitem[Froeschl\'{e} and Scholl(1989)]{Froeschle_1989} Froeschl\'{e}, 
Ch, Scholl, H., 1989, CMDA 46, 231.

\bibitem[Ivezi\'{c} et al.(2001)]{Ivezic_2001} Ivezi\'{c}, \v{Z}, and 
34 co-authors, 2001, AJ, 122, 2749.

\bibitem[Kne\v{z}evi\'{c} and Milani(2000)]{Knezevic_2000} Kne\v{z}evi\'{c}, 
Z., Milani, A. (2000), CMDA, 78, 17.

\bibitem[Masiero et al.(2012)]{Masiero_2012} Masiero, J. R., 
Mainzer, A. K., Grav, T., Bauer, J. M., and Jedicke, R., 2012, APJ, 759, 14. 

\bibitem[Morbidelli \& Vokrouhlick\'{y}(2003)]{Morby_2003} Morbidelli A., 
Vokrouhlick\'{y}, D., 2003, Icarus 163, 120.

\bibitem[Murray and Dermott(1999)]{Murray_1999} Murray, C. D., 
Dermott, S. F., 1999, Solar System Dynamics, Cambridge Univ. Press, Cambridge.

\bibitem[Nesvorn\'{y} et al.(2015)]{Nesvorny_2015} 
Nesvorn\'{y}, D., Bro\v{z}, M., Carruba, V. 2015,   
In Asteroid IV, (P. Michel, F. E. DeMeo, W. Bottke Eds.), Univ. Arizona Press 
and LPI, 297.

\bibitem[Novakovi{\'c} et al.(2010)]{Novakovic_2010} Novakovi{\'c}, B., 
Tsiganis, K., \& Kne{\v z}evi{\'c}, Z.\ 2010, MNRAS, 402, 1263.

\bibitem[Novakovi\'{c} et al.(2015)]{Novakovic_2015}
Novakovi\'{c}, B., Maurel, C., Tsirvoulis, G., Kne\v{z}evi\'{c}, Z.
2015, ApJ, 807, L5.

\bibitem[Novakovic et al.(2016)]{Novakovic_2016} Novakovic, B., 
Tsirvoulis, G., Maro, S., Djosovic, V., \& Maurel, C.\ 2016, arXiv:1601.00905 

\bibitem[Spoto et al.(2015)]{Spoto_2015} Spoto, F., 
Milani, A. Kne\v{z}evi\'{c}, Z. 2015, Icarus, 257, 275.

\bibitem[Tsirvoulis and Novakovi\'c(2016)]{Tsirvoulis_2016} Tsirvoulis, G,
Novakovi\'{c}, B. (2016) Icarus, in press.

\bibitem[Vokrouhlick\'{y} et al. (2006a)]{Vokrouhlicky_2006a} 
Vokrouhlick{\'y}, D., Bro{\v z}, M., Morbidelli, A., 
et al.\ 2006a, Icarus, 182, 92. 

\bibitem[Vokrouhlick\'{y} et al. (2006b)]{Vokrouhlicky_2006b} 
Vokrouhlick\'{y} D., Bro\v{z}, M., Bottke, W. F., Nesvorn\'{y}, D., 
Morbidelli, A. 2006b, Icarus, 182, 118.

\bibitem[Vokrouhlick\'{y} et al. (2006c)]{Vokrouhlicky_2006c} 
Vokrouhlick\'{y} D., Bro\v{z}, M., Bottke, W. F., Nesvorn\'{y}, D., 
Morbidelli, A. 2006c, Icarus, 183, 349.

\end{thebibliography}
\end{document}